\documentclass[prd,amsmath,showpacs,preprintnumbers,superscriptaddress,floatfix,twocolumn]{revtex4}
\usepackage{amsmath}
\usepackage{graphicx}

\newcommand{\oa}[1]{${\cal O}(a^{#1})$}

\newcommand{\gambar}{\overline{\gamma}}

\newcommand{\beqa}{\begin{eqnarray}}
\newcommand{\eeqa}{\end{eqnarray}}

\begin{document}

\preprint{\vbox{
\rightline{ADP-05-11/T621}
\rightline{}}}

\title{Scaling behavior of quark propagator in full QCD}
\author{Maria B.~Parappilly$^{1}$, Patrick O.\ Bowman$^{2}$, 
Urs M.\ Heller$^{3}$,
Derek B.\ Leinweber$^{1}$, Anthony G.\ Williams$^{1}$ and J.\ B.\ Zhang}
\affiliation{
{Special Research Center for the Subatomic Structure of
Matter (CSSM) and Department of Physics,
University of Adelaide 5005, Australia} \\
$^{2}${Nuclear Theory Center, Indiana University, Bloomington IN 47405,
USA} \\
$^{3}${American Physical Society, One
Research Road, Box 9000, Ridge, NY 11961-9000, USA}
}

\date{\today}

\begin{abstract} 
We study the scaling behavior of the quark propagator on two lattices with 
similar physical volume in Landau gauge with 2+1 flavors of dynamical quarks
in order to test whether we are close to the continuum limit for these lattices.  
We use configurations generated with an improved staggered (``Asqtad'') action 
by the MILC collaboration.  
The calculations are performed on $28^3\times 96$ lattices with lattice 
spacing $a  = 0.09$ fm and on $20^3\times 64$ lattices with lattice 
spacing $a  = 0.12$ fm.  We calculate the quark mass function, $M(q^2)$, 
and the wave-function renormalization function, $Z(q^2)$, for a variety of 
bare quark masses.  Comparing the behavior of these functions on the two sets
of lattices we find that both $Z(q^2)$ and $M(q^2)$ show little sensitivity to
the ultraviolet cutoff.
\end{abstract}


\maketitle

\section{Introduction}
Quantum Chromodynamics (QCD) is widely considered to be the correct theory of 
the strong interactions.  Quarks and gluons are the fundamental degrees of 
freedom of this theory. 
The quark propagator contains valuable information about nonperturbative QCD. 
The systematic study of the quark propagator on the lattice has provided 
fruitful interaction with other approaches to hadron physics, such as 
instanton phenomenology~\cite{Diakonov:2002fq}, chiral quark 
models~\cite{Arriola:2003bs} and Dyson-Schwinger equation 
studies~\cite{Bhagwat:2003vw,Alkofer:2003jj}. 
As a first principles approach lattice QCD has provided valuable constraints 
for model builders.  In turn, such alternative methods can provide feedback on 
regions that are difficult to access directly on the lattice, such as
the deep infrared and chiral limits.

The quark propagator has previously been studied using 
Clover~\cite{Skullerud:2000un,Skullerud:2001aw}, 
staggered~\cite{Bowman:2002bm,Bowman:2002kn} and 
Overlap~\cite{Bonnet:2002ih,Zhang:2003fa,Boucaud:2003dx} actions.  For a 
review, see Ref.~\cite{Bowman:Springer}.
All these actions have different systematic errors and the combination of 
these studies has given us an excellent handle on the possible lattice 
artifacts in quenched QCD.

In this study we focus on the Landau gauge quark propagator in full QCD, and 
extend our previous work~\cite{Bowman:2005vx} to a finer lattice 
with lattice spacing $a = 0.09$ fm~\cite{Aubin:2004wf} but similar physical volume in order to 
test whether we are close to the continuum limit for these lattices. The scaling 
behavior of the momentum space quark propagator is examined by comparing the 
results on these two lattices. 
Our results show that there are no significant differences in the wave-function 
renormalization function and quark mass function on the
two sets of lattices. Therefore the scaling behavior is good already at the coarser 
lattice spacing of $a = 0.12$ fm.

The configurations we use in this study were generated by the MILC 
collaboration~\cite{Aubin:2004wf,Bernard:2001av} and are available from the Gauge 
Connection~\cite{nersc}. The dynamical configurations have two degenerate light fermions for the $u$ 
and $d$ quarks and a heavier one for the strange quark. Weighting 
for the fermion determinants is provided by the so-called, ``fourth root 
trick.''. While the current numerical results~\cite{Durr:2005ax} provide
compelling evidence that the fourth root trick gives an accurate
estimate of the dynamical fermion weight, the formal issue of proving that this
provides the determinant of a local fermion action
from first principles remains unresolved.

\section{Details of the calculation}

The quark propagator is gauge dependent and we work in the Landau gauge
for ease of comparison with other studies.  Landau gauge is
a smooth gauge that preserves the Lorentz invariance of the theory, so
it is a popular choice. As derived in Ref.~\cite{Bonnet:1999mj} an improved
Landau-gauge-fixing functional, ${\cal F}_{\text{Imp}}^G \equiv \frac{4}{3}
{\cal F}_1^G - \frac{1}{12 u_0} {\cal F}_2^G$ is used
 where 
\begin{equation}
{\cal F}^{G}_{1}[\{U\}] = \sum_{\mu, x}\frac{1}{2} \mbox{Tr} \, \left\{
U^{G}_{\mu}(x) + U^{G}_{\mu}(x)^{\dagger} \right\},
\end{equation}
\begin{equation}
{\cal F}^{G}_2 = \sum_{x,\mu} \frac{1}{2} \mbox{Tr} \left\{
U^{G}_{\mu}(x) U^{G}_{\mu}(x+\hat{\mu}) + \mbox{h.c.} \right\}.
\end{equation}

\begin{equation}
U^{G}_{\mu}(x) = G(x) U_{\mu}(x) G(x+\hat{\mu})^{\dagger},
\end{equation}

\begin{equation}
G(x) = \exp \left\{ -i \sum_a \omega^a(x) T^a \right\},
\end{equation}
and $u_0$ is the plaquette measure of the mean link.
We adopt a ``steepest decents'' approach. The functional derivative of 
${\cal F}_{\text{Imp}}^G$ with respect to $\omega^a$ provide

\begin{equation}
\Delta_1(x) \equiv  \frac{1}{u_0} \sum_\mu \left [ U_\mu(x-\mu) 
		- U_\mu(x) - \mbox{h.c.} 
 		\right ]_{\text{traceless}} 
\end{equation}	     

\begin{equation}
\begin{split}
\Delta_2(x)  \equiv  \frac{1}{u_0^2} \sum_\mu \bigg[ 
		U_\mu(x-2\mu)U_\mu(x-\mu)\\-U_\mu(x)U_\mu(x+\mu)- \mbox{h.c.}\bigg]_{\text{traceless}} 
\end{split}   
\end{equation}
and
\begin{equation}
\Delta_{\text{Imp}}(x) \equiv  \frac{4}{3} \Delta_1(x) 
			- \frac{1}{12} \Delta_2(x) .
\end{equation}				  
The resulting gauge transformation is 
\begin{equation}
G_{\text{Imp}}(x)  = \exp \left\{ \frac{\alpha}{2} \Delta_{\text{Imp}}(x) \right\},
\label{gauge_trans}
\end{equation}	
where $\alpha$ is a tuneable step-size parameter. The gauge fixing algorithm proceeds by calculating
the relevant $\Delta_i$ in terms of the mean-field-improved links, and then
applying the associated gauge transformation, Eq.~(\ref{gauge_trans}),
to the gauge field.  The algorithm using conjugate gradient Fourier acceleration
is implemented in parallel, updating all 
links simultaneously, and is iterated until the Lattice Landau gauge condition
\begin{equation}
\theta_{\text{Imp}} = \frac{1}{V N_c} \sum_x \mbox{Tr} \left\{ \Delta_ {\text{Imp}}(x)
\Delta_{\text{Imp}}(x)^{\dagger}
\right\}
\end{equation}
is satisfied with accuracy of $\theta_i < 10^{-12}$.

As this gauge fixing finds a local minimum of the gauge fixing functional, we are 
necessarily sampling from the first Gribov region.  Our ensemble contains no 
gauge-equivalent configurations and hence has no Gribov copies as such.  However, our 
configurations are local minima and {\em absolute} minima and therefore are 
not from the Fundamental Modular Region ~\cite{Williams:2002dw}. It is known from previous 
$SU(3)$ studies that neither the gluon nor quark propagator display any obvious Gribov noise 
above and beyond the ensemble statistical noise and so we do not consider 
it further here~\cite{Giusti:2001kr,Sternbeck:2004qk,Sternbeck:2004xr}. It will 
be interesting to repeat this calculation for the Gribov-copy free Laplacian gauge, and
to do a systematic search for Gribov noise in Landau gauge , but these are left for future studies.

The MILC configurations were generated with the \oa{2} one-loop
Symanzik-improved L\"{u}scher--Weisz gauge action~\cite{Luscher:1984xn}.
The dynamical configurations use the Asqtad quark action~\cite{Asqtad}, 
an \oa{2} 
Symanzik-improved staggered fermion action which removes lattice artifacts 
up to order $a^2g^2$.  We refer to the $a = 0.09$ fm lattice as the ``fine''
lattice and the $a = 0.12$ fm one as the ``coarse'' lattice.
  
We explore two light sea quark masses, $ma = 0.0062$ ($m = 14.0$ MeV) 
and $ma = 0.0124$ ($m = 27.1$ MeV).  The bare strange quark mass was fixed at 
$ma = 0.031$, or $m = 67.8$ MeV for $a = 0.09$ fm.
The values of the coupling and the bare light sea-quark masses are matched such
that the lattice spacing is held constant.  The simulation parameters are 
summarized in Table~\ref{simultab} with the lattice spacings taken from
~\cite{Aubin:2004wf}.

\begin{table}[t!]
\caption{\label{simultab} Lattice parameters used in this study. The
dynamical configurations each have two degenerate light quarks
(up/down) and a heavier quark (strange). The light bare quark masses for the $28^3\times 96$ 
lattice are $14.0$ MeV and $27.1$ MeV 
with a strange quark mass of $67.8$ MeV. For the $20^3\times 64$ lattice the bare
quark masses range from $15.7$ MeV to $78.9$ MeV. The lattice spacing
is $a \simeq 0.12$ fm  for the $20^3\times 64$ lattice 
and $a \simeq 0.09$ fm ~\cite{Aubin:2004wf} for the $28^3\times 96$ lattice.}    
\begin{ruledtabular}
\begin{tabular}{ccccccccc}
   & Dimensions      & $\beta$ & $a$       &Bare Quark Mass  & \# Config \\   
\hline
1  & $28^3\times 96$ &   7.09  & 0.086 fm  &$14.0$ MeV, $67.8$ MeV  & 108 \\
2  & $28^3\times 96$ &   7.11  & 0.086 fm  &$27.1$ MeV, $67.8$ MeV  & 110 \\
\hline
3  & $20^3\times 64$ &   6.76  & 0.121 fm  &$15.7$ MeV, $78.9$ MeV  & 203 \\ 
4  & $20^3\times 64$ &   6.79  & 0.121 fm  &$31.5$ MeV, $78.9$ MeV  & 249 \\ 
5  & $20^3\times 64$ &   6.81  & 0.120 fm  &$47.3$ MeV, $78.9$ MeV  & 268 \\
6  & $20^3\times 64$ &   6.83  & 0.119 fm  &$63.1$ MeV, $78.9$ MeV  & 318 
\end{tabular}
\end{ruledtabular}
\end{table}

\begin{figure}[t]
\centering\includegraphics[height=0.99\hsize,angle=90]{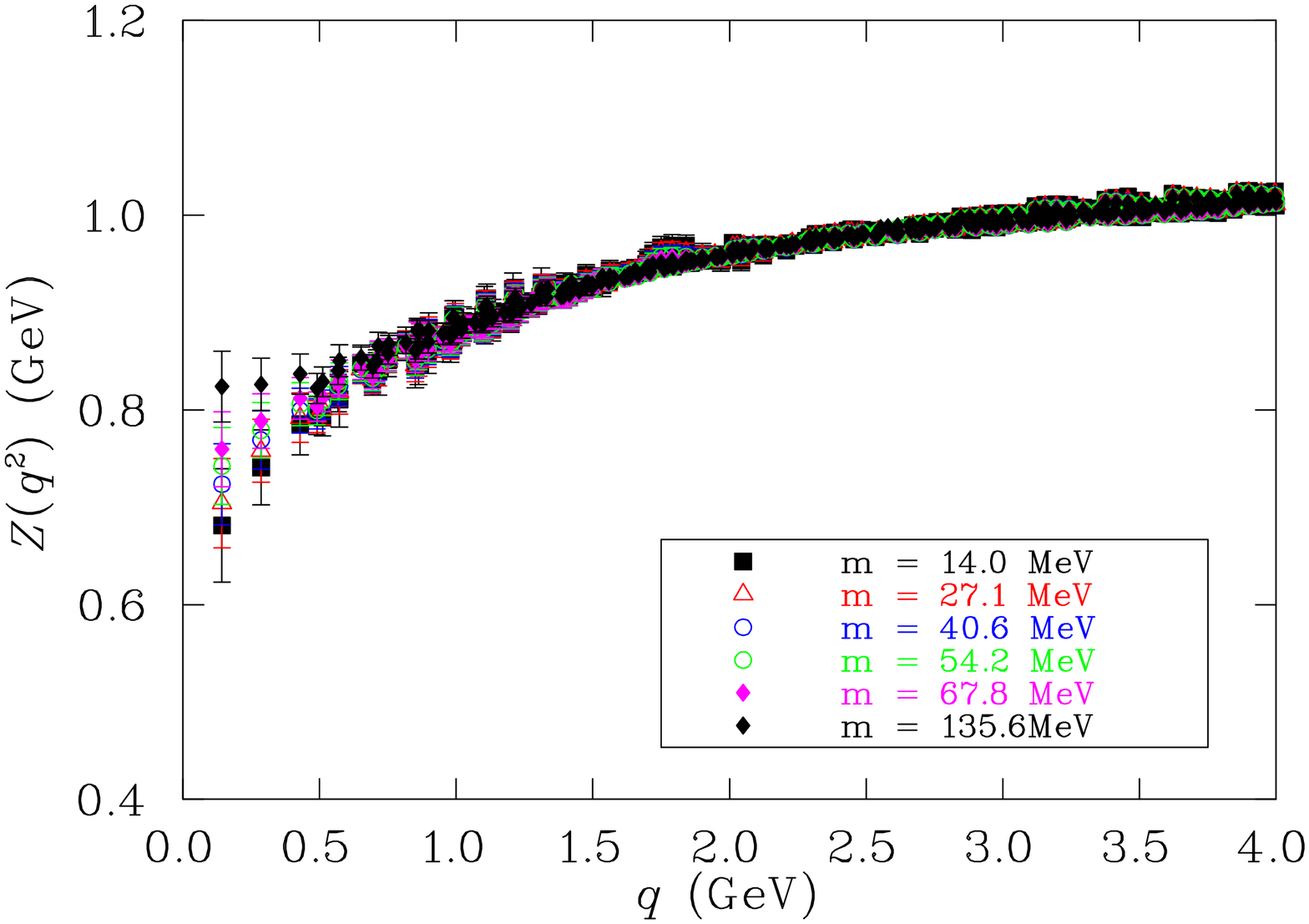}

\vspace*{5mm}

\centering\includegraphics[height=0.99\hsize,angle=90]{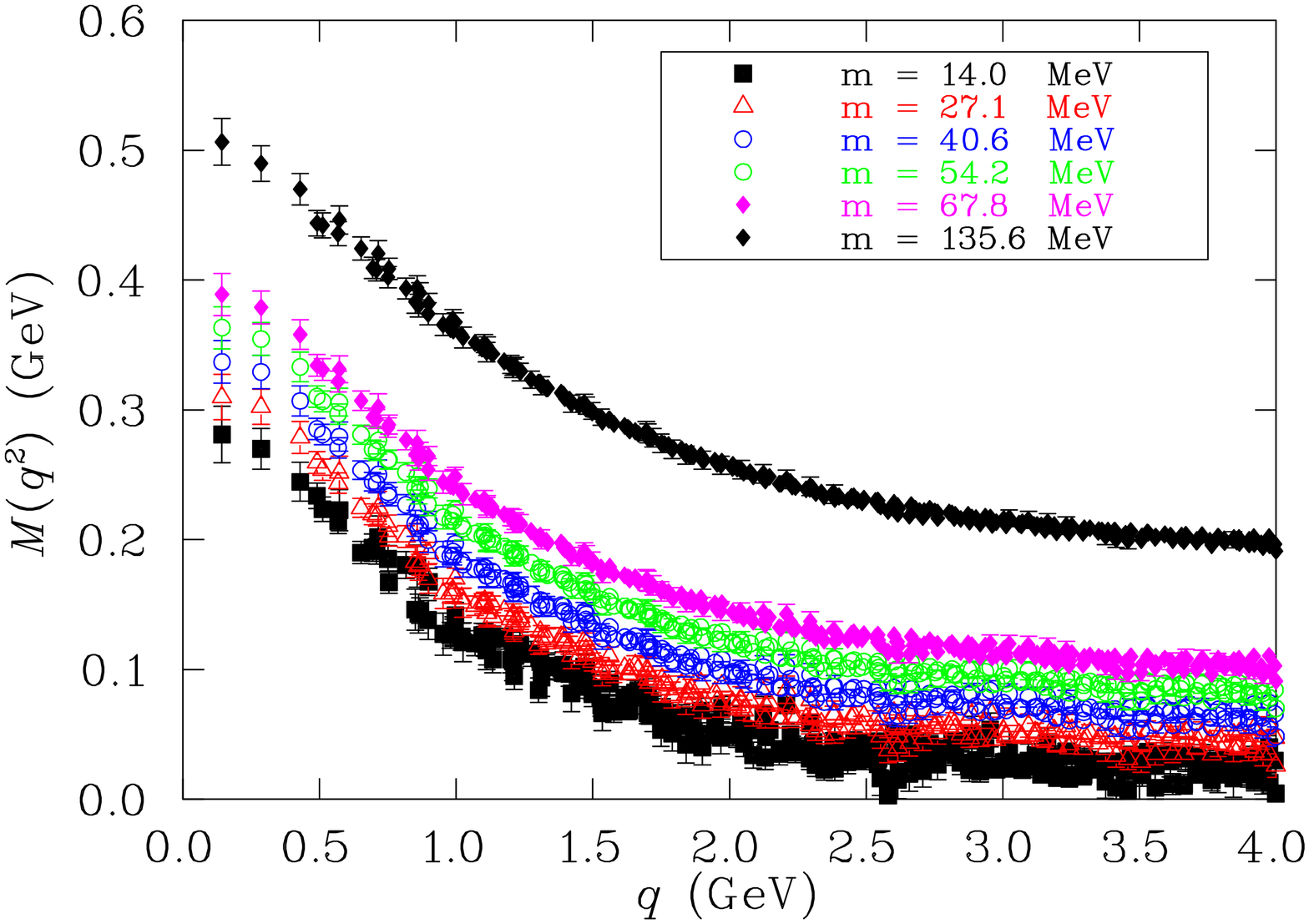}
\caption{The unquenched wave-function renormalization function $Z(q^2)$ and 
mass function $M(q^2)$ for a  variety of valence quark masses (shown in the 
inset), with the light sea-quark mass fixed at $m = 14.0$ MeV. The 
renormalization function is renormalized at $q$ = $3.0$ GeV.}
\label{smallmass}
\end{figure}

On the lattice, the bare propagator $S(a; p^2)$ is related to the
renormalized propagator $ S^{\text{ren}}(\mu;p^2)$ through the
renormalization constant~\cite{Bowman:2005vx}
\begin{equation}
S(a;p^2) = Z_2(a;\mu) S^{\text{ren}}(\mu;p^2).
\end{equation}
In the continuum limit, Lorentz invariance allows one to decompose
the full quark propagator into Dirac vector and scalar pieces
\begin{equation}
S^{-1}(p^2) = Z^{-1}(p^2) [i \gamma \cdot p + M(p^2)],
\end{equation}
where $M(p^2)$ and $Z(p^2)$ are the nonperturbative mass and wave function
renormalization functions, respectively.  Asymptotic freedom implies
that, as $p^2 \rightarrow \infty$, $S(p^2)$ reduces to the tree-level
propagator
\begin{equation}
\label{eq:free_quark}
S^{-1}(p^2) \rightarrow  i\gamma \cdot p + m,
\end{equation}
up to logarithmic corrections.  The mass function $M$ is renormalization point 
independent and for $Z$ we choose throughout this work the renormalization 
point as $\mu= 3.0$ GeV, {\em i.e.},

\begin{equation}
S^{\text{ren}}(\mu;\mu^2) = \frac{S(a;\mu^2)}{Z_2(a;\mu)}=  1,
\end{equation}
thus defining $Z_2(a;\mu)$.

The tree-level quark propagator with the Asqtad action has the form
\begin{equation}
S^{-1}(p) = i \sum_\mu \gambar_\mu q(p_\mu) + m,
\end{equation}
where $q(p_\mu)$ is the kinematic momentum given in~\cite{Bowman:2002bm}
\begin{equation}
\label{eq:mom_Asqtad}
q_\mu \equiv \sin(p_\mu) \bigl[ 1 + \frac{1}{6} \sin^2(p_\mu) \bigr].
\end{equation}
The $\gambar_\mu$ form a staggered Dirac algebra (see Eq.(A.6) and (A.7) of 
Ref.~\cite{Bowman:2005vx}).
Having identified the kinematic momentum, we define the mass and
renormalization functions by
\begin{equation}
S^{-1}(p)  = Z^{-1}(q) \Bigl [ i \sum_\mu {(\gambar_\mu)}
     q_\mu(p_\mu) + M(q) \Bigr].
\end{equation}
Additional details can be found in Ref.~\cite{Bowman:2005vx}

\section{Numerical Results}

\begin{figure}[t]
\centering\includegraphics[height=0.99\hsize,angle=90]{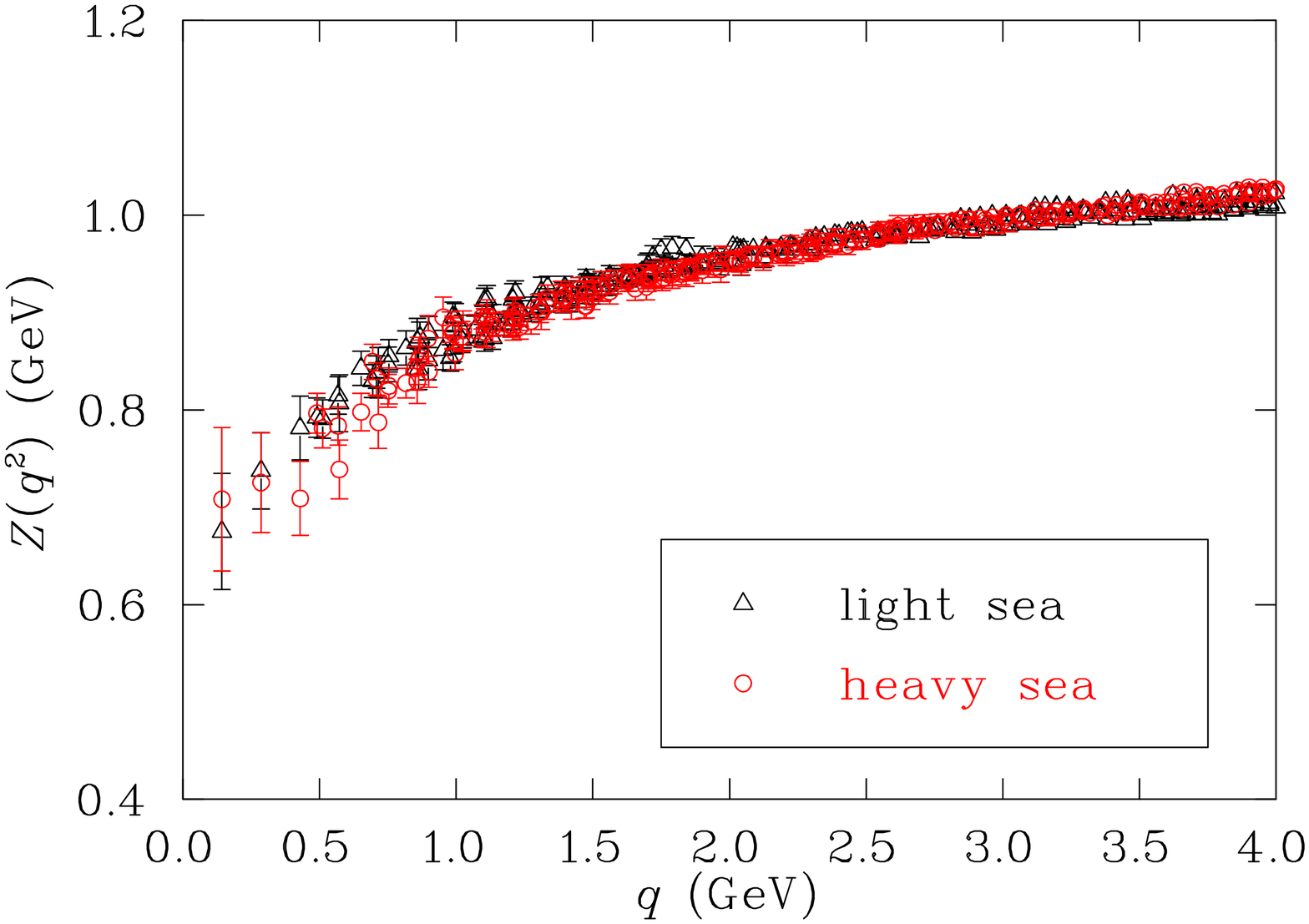}
\vspace*{5mm}
\centering\includegraphics[height=0.99\hsize,angle=90]{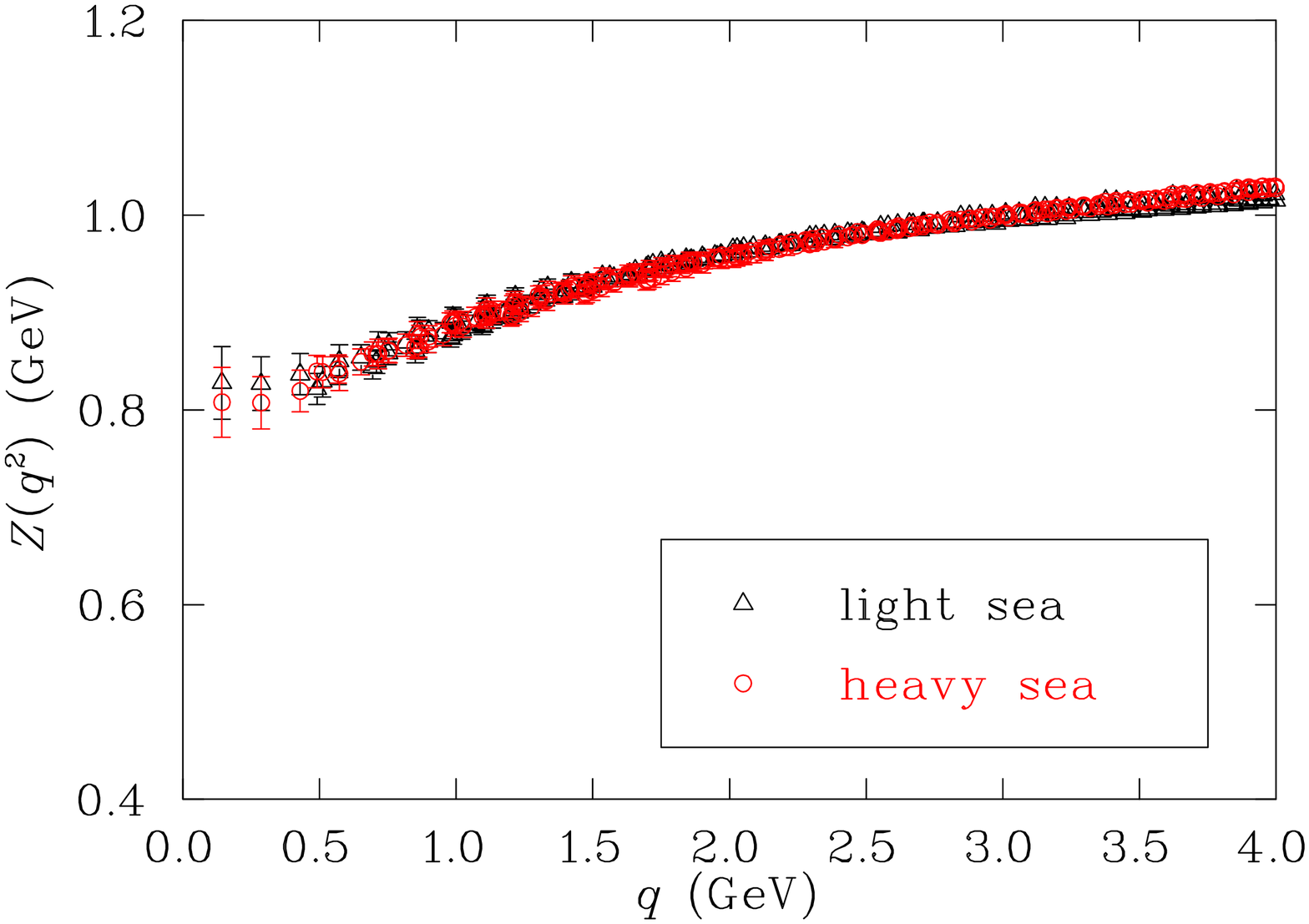}
\caption{The unquenched wave-function renormalisation function for the two 
different values of the light sea quark mass on the fine lattice 
($14.0$ MeV and $27.1$ MeV).  The valence quark masses are $m = 14.0$ MeV 
(top) and  $m = 135.6$ MeV (bottom), the lightest and heaviest in our 
current sample respectively. The renormalization function is renormalized at 
$q$ = $3.0$ GeV.}
\label{Zlightandheavy}
\end{figure}

\begin{figure}[b]
\centering\includegraphics[height=1.105\hsize,angle=90]{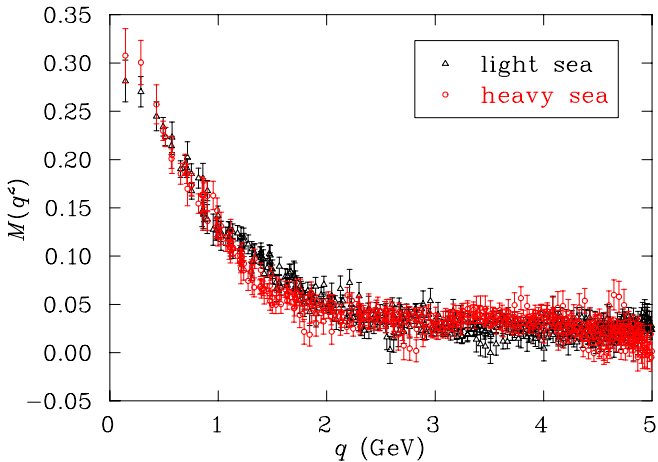}
\vspace*{-2.5mm}
\centering\includegraphics[height=1.105\hsize,angle=90]{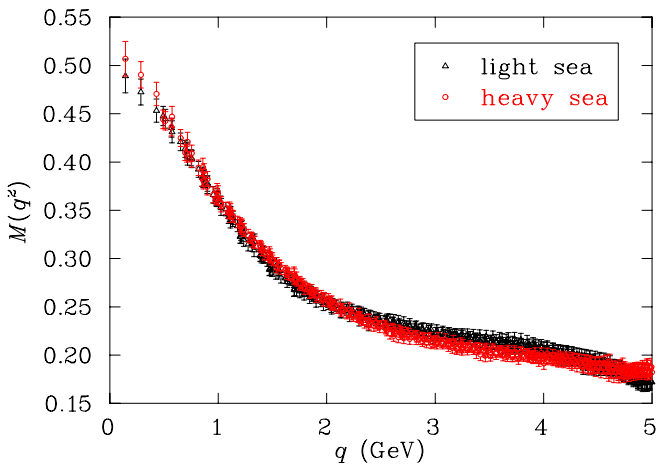}
\vspace*{-7mm}
\caption{The unquenched quark mass function for the two different values of the
light sea quark mass on the fine lattice ($14.0$ MeV and $27.1$ MeV). The valence quark masses are 
$m = 14.0$ MeV (top) and  $m = 135.6.$ MeV (bottom), the lightest and heaviest in our current sample 
respectively.}
\label{Mlightandheavy}
\end{figure}

In Fig.~\ref{smallmass} we show the results for the mass function $M(q^2)$ 
and wave-function renormalization function $Z(q^2)$ for the lightest of our 
light sea quark masses for a variety of valence quark masses.
In these figures, one valence quark mass ($14.0$ MeV) is identical to the light sea quark 
mass, as in full QCD.  The others are partially quenched results. 
The data are ordered as we expect, i.e., the larger the bare valence quark
mass, the higher $M(q^2)$.  The wave-function renormalization function, 
$Z(q^2)$, on the other hand, is infrared suppressed and the smaller the
valence quark mass, the more pronounced the dip at low momenta.
In Figs.~\ref{Zlightandheavy} and \ref{Mlightandheavy} we instead hold 
the valence quark mass fixed and vary the
sea quark mass.  Clearly the dependence over this small range of sea quark 
masses is weak.  Unfortunately we only have two dynamical sets to compare, and
for the lightest valence quark the data are rather noisy.

Next we work on two lattices with different lattice spacing but similar 
physical volume.  We compare the wave-function renormalization function $Z(q^2)$ and 
mass function $M(q^2)$ for two lattices with different lattice spacing $a$ in 
full lattice QCD.

\begin{figure}[t]      
\centering\includegraphics[height=0.99\hsize,angle=90]{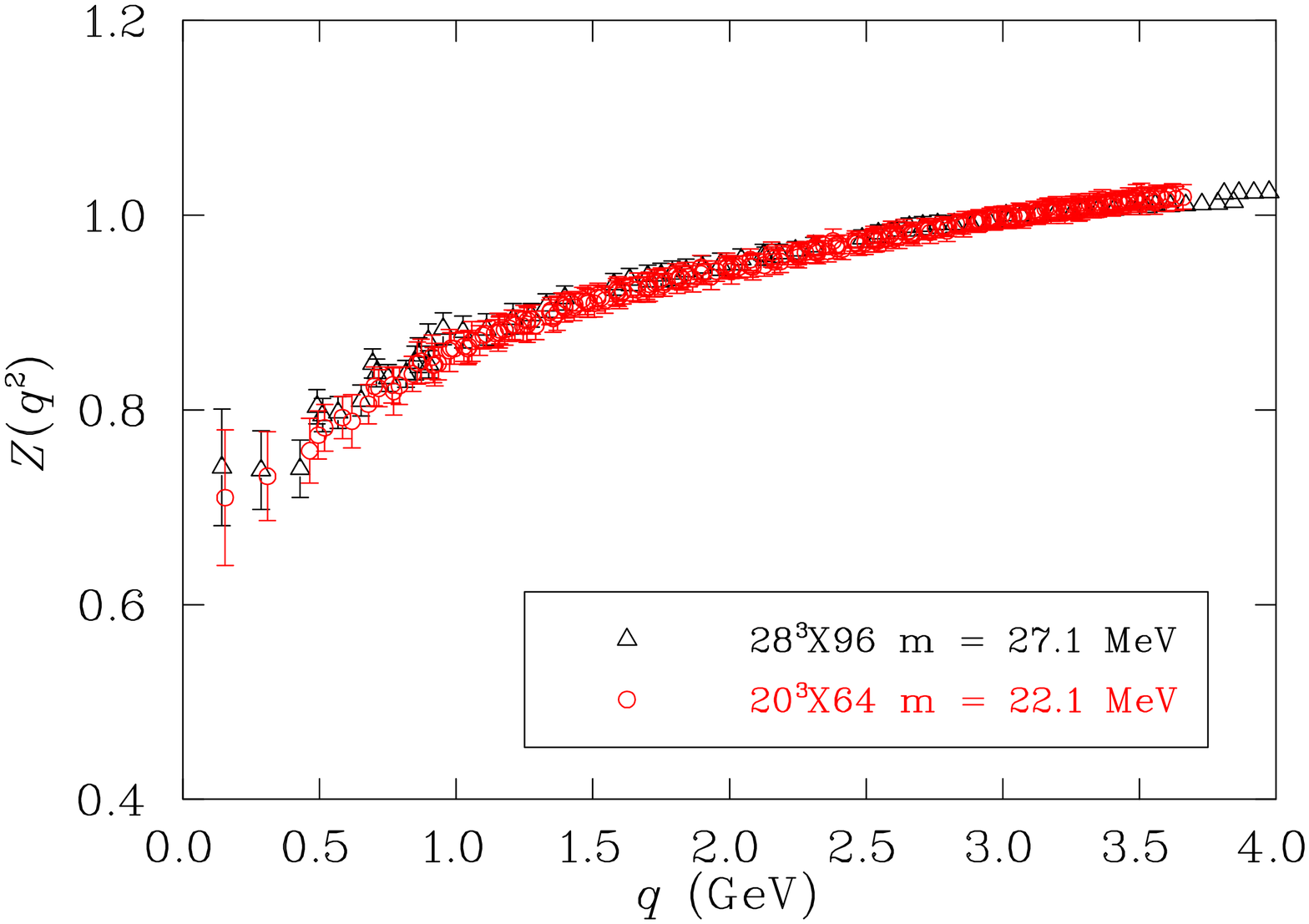}
\vspace*{5mm}
\centering\includegraphics[height=0.99\hsize,angle=90]{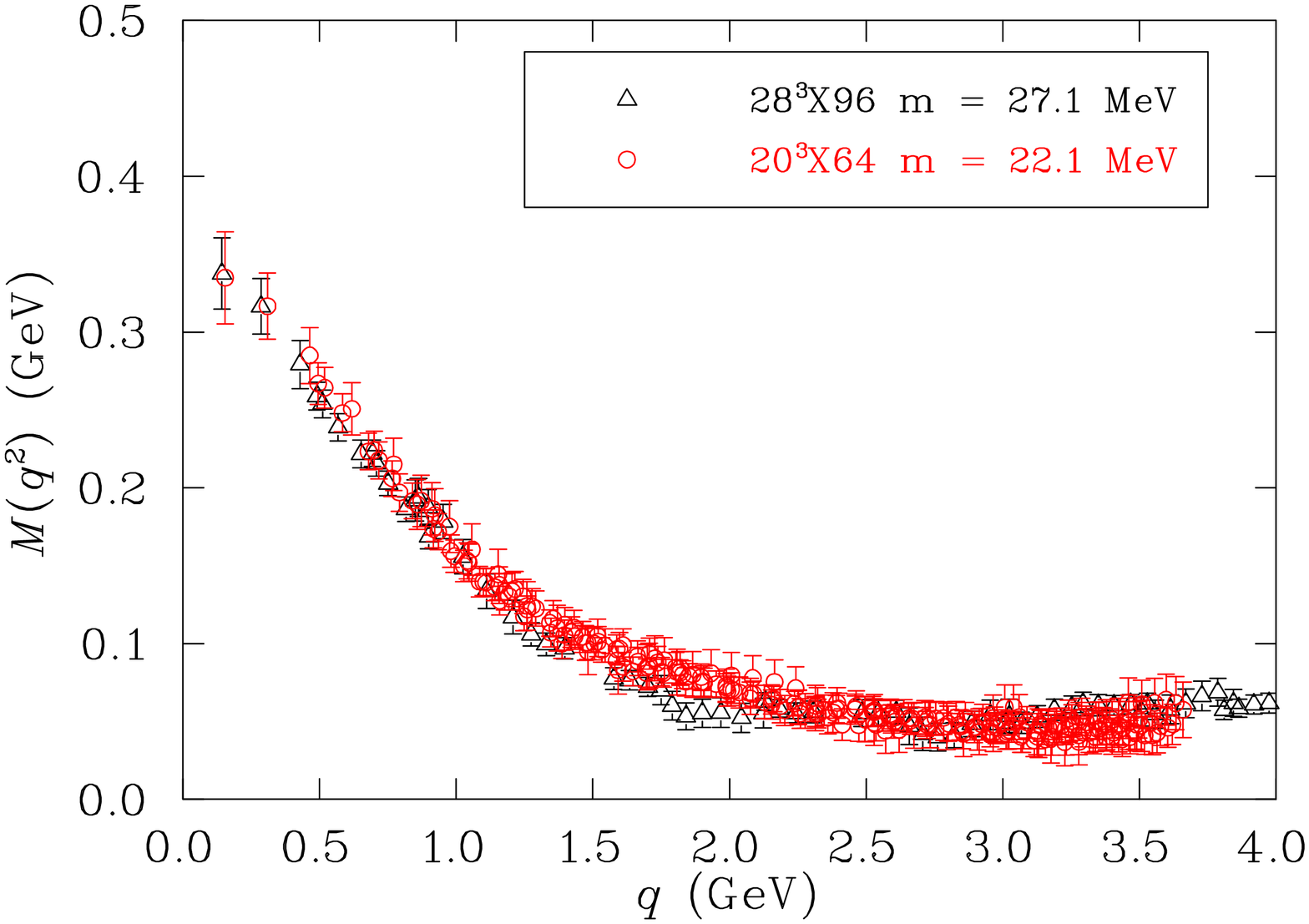}
\caption{Comparison of wave-function renormalization function $Z(q^2)$ and 
mass function $M(q^2)$ for two different lattices. 
Triangles correspond to the quark propagator at  mass $m = 27.1$ MeV from 
$28^3\times 96$ with lattice spacing $a = 0.09$ fm.
The open circles are the data from $20^3\times 64$ with lattice spacing 
$a = 0.12$ fm obtained by interpolating four different set of light quark 
masses making the $M(q^2)$ value matched for both lattices at  $q$ = $3.0$ 
GeV.  The renormalization point for $Z(q^2)$ is set at $q$ = $3.0$ GeV for 
both lattices.}
\label{interpolbig}
\end{figure}

\begin{figure}[t]      
\centering\includegraphics[height=0.99\hsize,angle=90]{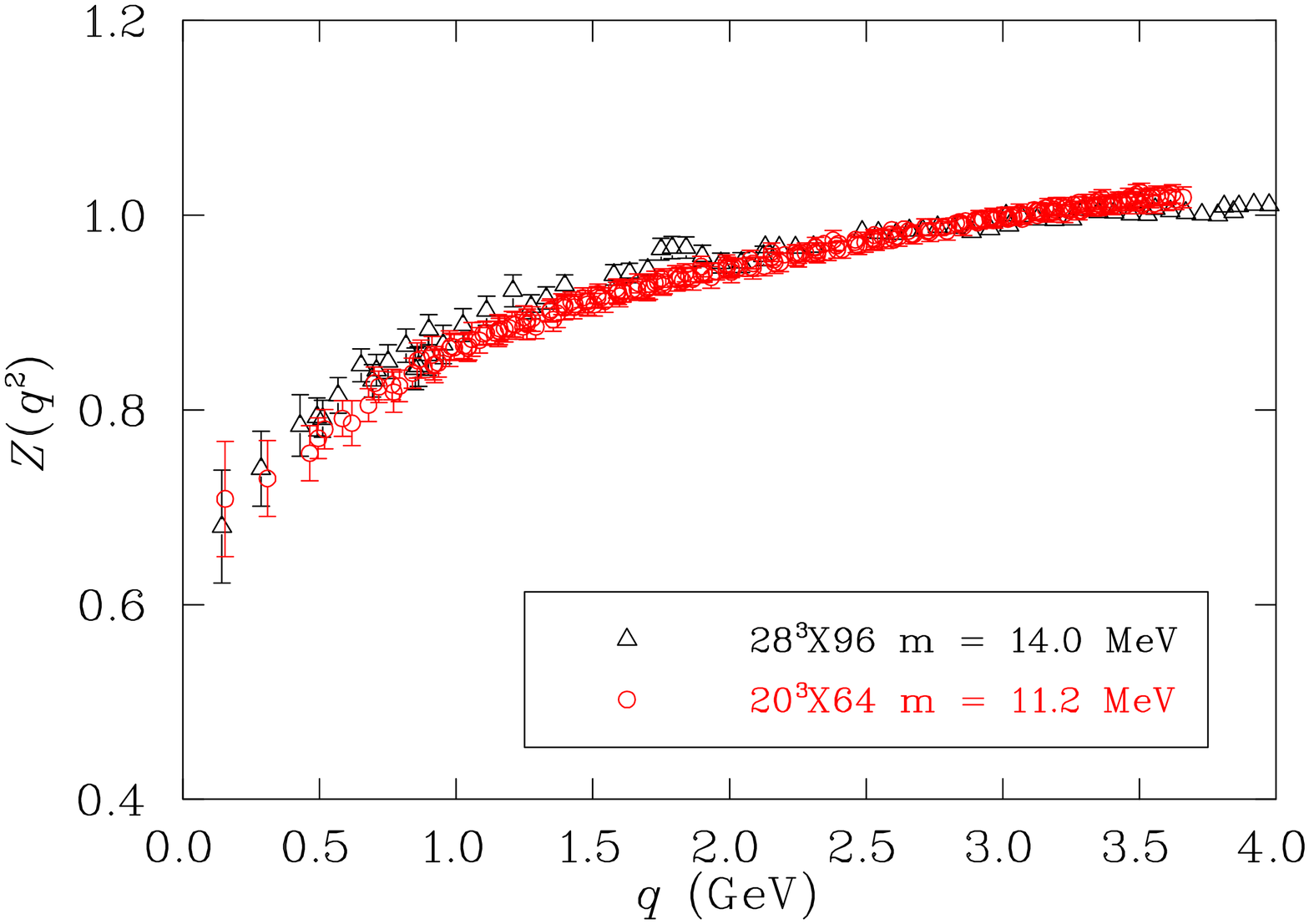}
\vspace*{5mm}
\centering\includegraphics[height=0.99\hsize,angle=90]{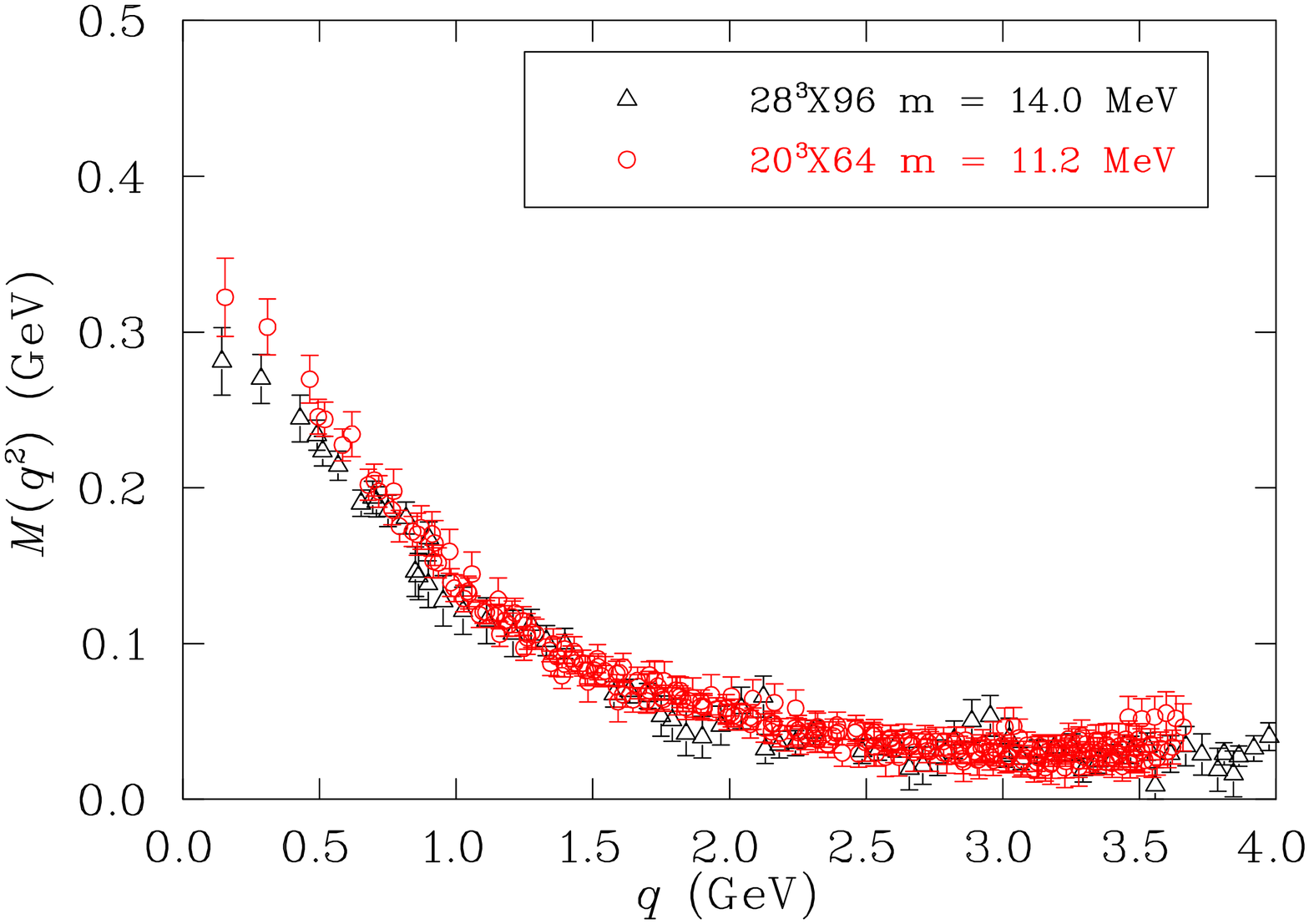}
\caption{This figure is same as figure 3, except the light quark mass of 
$28^3\times 96$ with lattice spacing $a= 0.09$ fm is $m = 14.0$ MeV.  The 
renormalization point for $Z(q^2)$ is set at $q$ = $3.0$ GeV for both 
lattices.}
\label{interpolsmall}
\end{figure}

In Fig.~\ref{interpolbig}, we show the quark propagator from the fine lattice 
for full QCD (light sea-quark mass and valence quark mass equal) with the light 
quark mass set to $m$ = $27.1$ MeV. This is compared with data from the 
coarse lattice by a simple linear interpolation from the four different data 
sets so the running masses are the same at $q^2$ = $3.0$ GeV.  
Fig.~\ref{interpolsmall} repeats this for the lighter sea quark, $m = 14.0$ MeV.
The quark propagators are in excellent agreement, showing no dependence on the 
lattice spacing.
\section{Conclusions}
In this study we performed a systematic comparison of the Asqtad quark 
propagator in full QCD for two lattices with different lattice spacing in order to
establish how close these lattices are to the scaling region and hence to
the contiuum limit. 
We compared the two functions $Z(q^2)$ and $M(q^2)$ on fine and coarse lattices
and found them to be consistent within errors.
We can thus deduce that for both lattices we are close to the scaling region
for the quark propagator, which for example makes these lattices suitable for 
future studies attempting to determine quark masses~\cite{Becirevic:1999kb}. 

\clearpage
\section*{ACKNOWLEDGMENTS}
We thank the Australian Partnership for Advanced Computing (APAC)
and the South Australian Partnership for Advanced Computing (SAPAC)
for generous grants of supercomputer time which have enabled this
project.  POB thanks the CSSM for its hospitality during part of this work.
This work is supported by the Australian Research Council.



\end{document}